\documentclass[final,3p,times]{elsarticle}

\usepackage{amssymb}

\usepackage{amsmath}
\usepackage{graphicx}
\usepackage{subcaption}
\usepackage{float}

\journal{Journal of Subatomic Particles and Cosmology}

\begin{document}

\begin{frontmatter}



\title{Quarkonium production in light-ion collisions with the ALICE experiment}

\author[aaa]{Rebecca Cerri, on behalf of the ALICE Collaboration}
\affiliation[aaa]{organization={University of Turin},
             addressline={Via Pietro Giuria 1},
             city={Turin},
             postcode={10125},
             country={Italy}}

\begin{abstract}
Quarkonium production has long been considered as one of the golden probes to study the quark–gluon plasma (QGP). In fact, the early production of heavy quarks ($\mathrm{c}\overline{\mathrm{c}}$ and $\mathrm{b}\overline{\mathrm{b}}$) makes quarkonia an ideal tool to investigate the evolution of the hot and dense medium produced in ultra-relativistic heavy-ion collisions. In such a medium, quarkonium production is expected to be suppressed due to the screening of the binding potential between the heavy quark and antiquark and/or through in-medium interactions. Moreover, at LHC energies the recombination of uncorrelated charm quarks pairs, namely regeneration, was found to significantly affect charmonium observables, in contraposition to the suppression mechanism. In addition, measurements in smaller collision systems as p–Pb have highlighted the possibility to observe QGP-like effects. In this context, the study of charmonium production in intermediate collision systems, as light-ion collisions, becomes more and more interesting, representing an ideal test ground for the state-of-the art theoretical models. In this contribution the new measurements of charmonium production  will be shown using the light-ion collisions data collected for the first time at the LHC in 2025 (oxygen-oxygen (OO), proton-oxygen (pO)). The results will be shown exploiting the forward ALICE rapidity coverage (2.5 < \textit{y} < 4). Finally, the measurements will be compared with the existing theoretical models.
\end{abstract}




\end{frontmatter}



\section{Introduction}
\label{sec1}
At ultra-relativistic energies, collisions of heavy nuclei produce a state of strongly interacting matter in which quarks and gluons are no longer confined inside hadrons, known as the quark-gluon plasma (QGP) \cite{ref:QGPBusza}. Heavy quarks, such as charm (c) and beauty (b), are important probes of this phase of matter because they are created in the initial hard partonic scatterings and undergo the full evolution of the system \cite{ref:HeavyQuark1}. In particular, the high-density of free color charges in the QGP leads to the dissociation of quarkonium states ($\mathrm{c}\overline{\mathrm{c}}$ or $\mathrm{b}\overline{\mathrm{b}}$) through screening of the quark--antiquark bond and/or in-medium interactions. The degree of dissociation depends on the binding energy of each state and on the temperature of the system \cite{ref:Suppression1}. On the other hand, quarkonia can also be produced through the recombination of (partially) thermalized heavy quarks, either during the QGP evolution~\cite{ref:recombination2} or at the QGP phase boundary~\cite{ref:recombination3}. When the number of heavy quarks is sufficiently large, recombination can partially counterbalance the suppression mechanism at $p_T \lesssim 5$ GeV/$c$. The modification of particle production in nucleus-nucleus (AA) collisions is quantified using the nuclear modification factor $R_{\rm AA}$ \cite{ref:Raa}, defined as the ratio between the production yield in AA collisions and that in proton-proton (pp) collisions, normalized to the number of nucleon-nucleon (NN) interactions. The ALICE Collaboration measured a suppression of $\mathrm{J}/\mathrm{\psi}$ production in Pb–Pb collisions, consistent with the expected dissociation of quarkonium states in the QGP \cite{ref:Raa_ALICE}. The $R_{\rm AA}$ reduction was found to be less pronounced than that observed at the SPS and RHIC energies, particularly at low transverse momentum. This observation is interpreted as evidence for a significant contribution from charm quark recombination at the LHC energies, where the larger charm production cross-section increases the probability of $\mathrm{J}/\mathrm{\psi}$ regeneration \cite{ ref:recombination2,ref:Raa_ALICE}. The study of the nuclear modification factor across collision systems of different sizes and shapes becomes particularly relevant, as it provides a connection between pp and Pb–Pb collisions. Previous measurements in p–Pb collisions have shown a suppression in $\mathrm{J}/\mathrm{\psi}$ $R_{\rm pPb}$ connected with the cold nuclear matter effects. This was found to be compatible with expectations from nuclear shadowing \cite{ref_ALICE_RpPb, ref:shadowing}, although the presence of additional or alternative cold nuclear matter effects, such as fully coherent energy loss (FCEL) \cite{Arleo_2013}, cannot be excluded. More recently, newly collected data in oxygen-oxygen (OO) and proton-oxygen (pO) allow the previous studies to be extended. These new colliding systems offer the possibility to study how quarkonium production evolves in light-ion collision at the same LHC energy and to investigate whether the observed modifications arise predominantly from cold nuclear matter effects or whether hot nuclear effects also play a role. The first $\mathrm{J}/\mathrm{\psi}$ measurements in OO and pO collisions provide an initial step toward addressing these questions.

\section{The ALICE experiment}
\label{sec1}
The ALICE experiment was upgraded during the LS2 and the current detector setup used in Run 3 is described in detail in Ref. \cite{ALICE_upgrades}. The main subdetectors used in this analysis are the forward muon spectrometer (MCH), the Fast Interaction Trigger (FIT) and the Inner Tracking System (ITS). The first, covering a pseudorapidity region of -4.0 $< \eta <$ -2.5, includes a 3~Tm dipole magnet, two hadron absorbers, five tracking stations, and two muon identification stations (MID). The FIT system consists of 5 separate detectors distributed along the beam axis from $-19.5$~m to $+17$~m around the nominal interaction point. Among its main functions, FIT provides the interaction trigger and measures the charged-particle multiplicity in the forward region. The ITS is the innermost detector in the barrel
enabling the precise determination of the primary interaction vertex (PV).

\section{Analysis}
\label{sec1}
This section reports a preliminary study of the $\mathrm{J}/\mathrm{\psi}$ nuclear modification factor in OO collisions at the center of mass energy $\sqrt{s_{\mathrm{NN}}}=5.36$ TeV and in pO collisions at $\sqrt{s_{\mathrm{NN}}}=9.62$ TeV, using data recorded by ALICE in July 2025. The data samples were collected during a short two-day data taking period, reaching recorded integrated luminosities of 5.01 nb$^{-1}$ and 7.27 nb$^{-1}$ for OO and pO collisions, respectively. The analysis focuses on the $\mathrm{J}/\psi$ decay into muon pairs within the rapidity range $2.5 < y < 4$, 
with results integrated over rapidity and centrality. In pO collisions, the two beams have different energies per nucleon, resulting in a shifted centre-of-mass frame. Consequently, the laboratory acceptance 
$2.5 < y_{\mathrm{lab}} < 4.0$ corresponds to $2.15 < y_{\mathrm{cms}} < 3.65$. Events are selected by requiring a coincidence of signals in the FIT detectors, while additional track quality selections are applied at the single muon and muon pair levels.
Muon tracks are reconstructed by combining information from the MCH and MID, and dimuons formed by muons of opposite charge are then selected for the analysis.
The nuclear modification factor in each tranverse momentum interval, $p_{\mathrm{T}}$, is calculated as:
\begin{equation}
\begin{aligned}
R_{\mathrm{OO}} &=
\frac{\sigma^{J/\psi}_{\mathrm{OO}}}
{A^{2}_{\mathrm{O}} \cdot \sigma^{J/\psi}_{\rm{pp}}},
\qquad
R_{\mathrm{pO}} &=
\frac{\sigma^{J/\psi}_{\mathrm{pO}}}
{A_{\mathrm{O}} \cdot \sigma^{J/\psi}_{\rm{pp}}},
\end{aligned}
\label{Eq:Method1}
\end{equation}
where $\sigma^{J/\psi}_{\mathrm{OO}}$ (or $\sigma^{J/\psi}_{\mathrm{pO}}$) is the J/$\psi$ cross section in OO (or pO) collisions, while $\sigma^{J/\psi}_{\rm{pp}}$ is the corresponding reference cross section in pp collisions. $A$ denotes the mass number, being $A_{\mathrm{O}}$ for pO collisions and $A_{\mathrm{O}}^2$ for OO collisions. This formulation, based on cross sections and mass numbers, is used since the results are obtained for the 0--100\% centrality interval. The $\sigma^{J/\psi}_{\mathrm{OO}}$ (or $\sigma^{J/\psi}_{\mathrm{pO}}$) is measured as:
\begin{equation}
\begin{aligned}
\sigma^{J/\psi}_{\mathrm{OO}} &=
\frac{N^{J/\psi}_{\mathrm{OO}}}
{\mathcal{L}_{\rm OO} \cdot (A \times \varepsilon)_{\rm OO} \cdot \mathrm{BR}},
\qquad
\sigma^{J/\psi}_{\mathrm{pO}} &=
\frac{N^{J/\psi}_{\mathrm{pO}}}
{\mathcal{L}_{\rm pO} \cdot (A \times \varepsilon)_{\rm pO} \cdot \mathrm{BR}}.
\end{aligned}
\label{Eq:Method1_bis}
\end{equation}
Here, $N^{J/\psi}_{\mathrm{OO}}$ (or $N^{J/\psi}_{\mathrm{pO}}$) is the number of reconstructed J/$\psi$ candidates in OO (or pO) collisions, $\mathcal{L}_{\rm OO} $ (or $\mathcal{L}_{\rm pO} $) the luminosity, $(A \times \varepsilon)_{\rm OO}$ (or $(A \times  \varepsilon)_{\rm pO}$) the product of acceptance and efficiency, and BR the $\mathrm{J}/\mathrm{\psi}$ branching ratio to dimuon decays. The number of reconstructed $\mathrm{J}/\mathrm{\psi}$ candidates is determined through likelihood fits to the opposite-sign dimuon invariant-mass spectra. The resonance signal is described with a double-sided Crystal Ball function, with the $\mathrm{J}/\mathrm{\psi}$ pole mass and width left as free parameters in the fit. The non-Gaussian tail parameters are constrained using either the same data sample or Monte Carlo simulations. The background below the signal is described using several empirical functions. The luminosity after selections has a value of 3.3 nb$^{-1}$ $\pm$ 3\% for OO, 6.3~nb$^{-1}$ $\pm$ 3.5\% for pO, which are used in Eq. \ref{Eq:Method1_bis}. The $ A \times \varepsilon$ is obtained using dedicated Monte Carlo simulation as a function of $p_{\mathrm{T}}$, with values ranging from 30\% to 47\% for both systems. In this analysis, the pp reference cross-section for both collision systems is obtained from an interpolation based on previously published ALICE measurements. The available J/$\psi$ cross-section measurements at $\sqrt{s}=2.76$, 5.02, 7, 8, and 13 TeV \cite{ALICE_13} are fitted with three different functions, following the procedure described in Ref. \cite{ALICE_interpolation}. The systematic uncertainties associated with the MCH--MID matching efficiency, the MCH tracking efficiency and the luminosity are evaluated by varying selection criteria. The dominant systematic uncertainty contribution in both analyses arises from the pp reference.

\begin{figure}[t]
    \centering

    \begin{minipage}[t]{0.38\textwidth}
        \centering
        \includegraphics[
            width=\linewidth,
            height=5.5cm,
            keepaspectratio
        ]{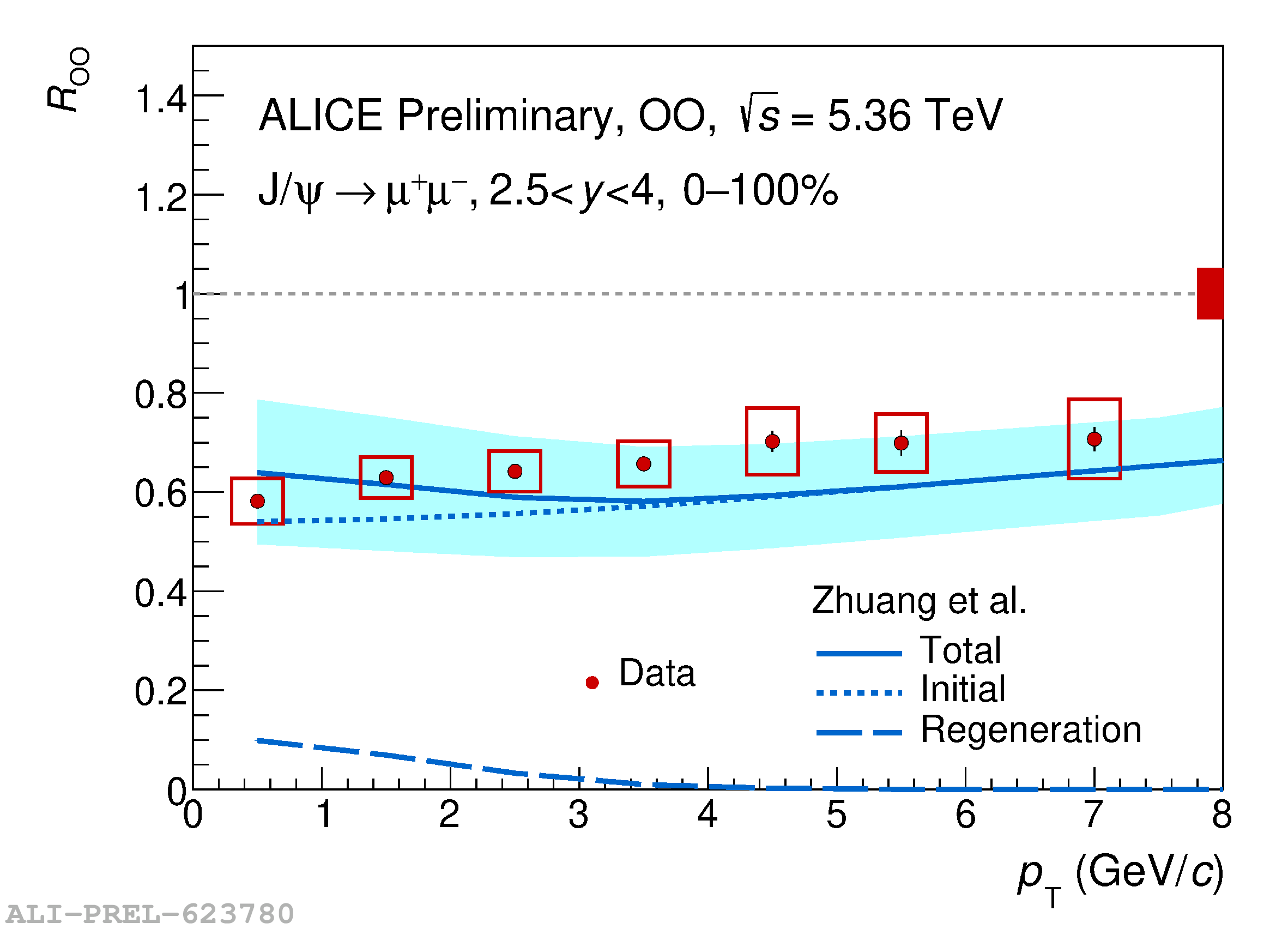}

        \caption{J/$\psi$ nuclear modification factor in OO collisions as a function of $p_{\mathrm{T}}$  at $\sqrt{s_{\mathrm{NN}}} = \text{5.36}$ TeV, compared with transport model prediction.}
        \label{fig:ROO_with_model}
    \end{minipage}
    \hspace{0.02\textwidth}
    \begin{minipage}[t]{0.40\textwidth}
        \centering
        \includegraphics[
            width=\linewidth,
            height=5.5cm,
            keepaspectratio
        ]{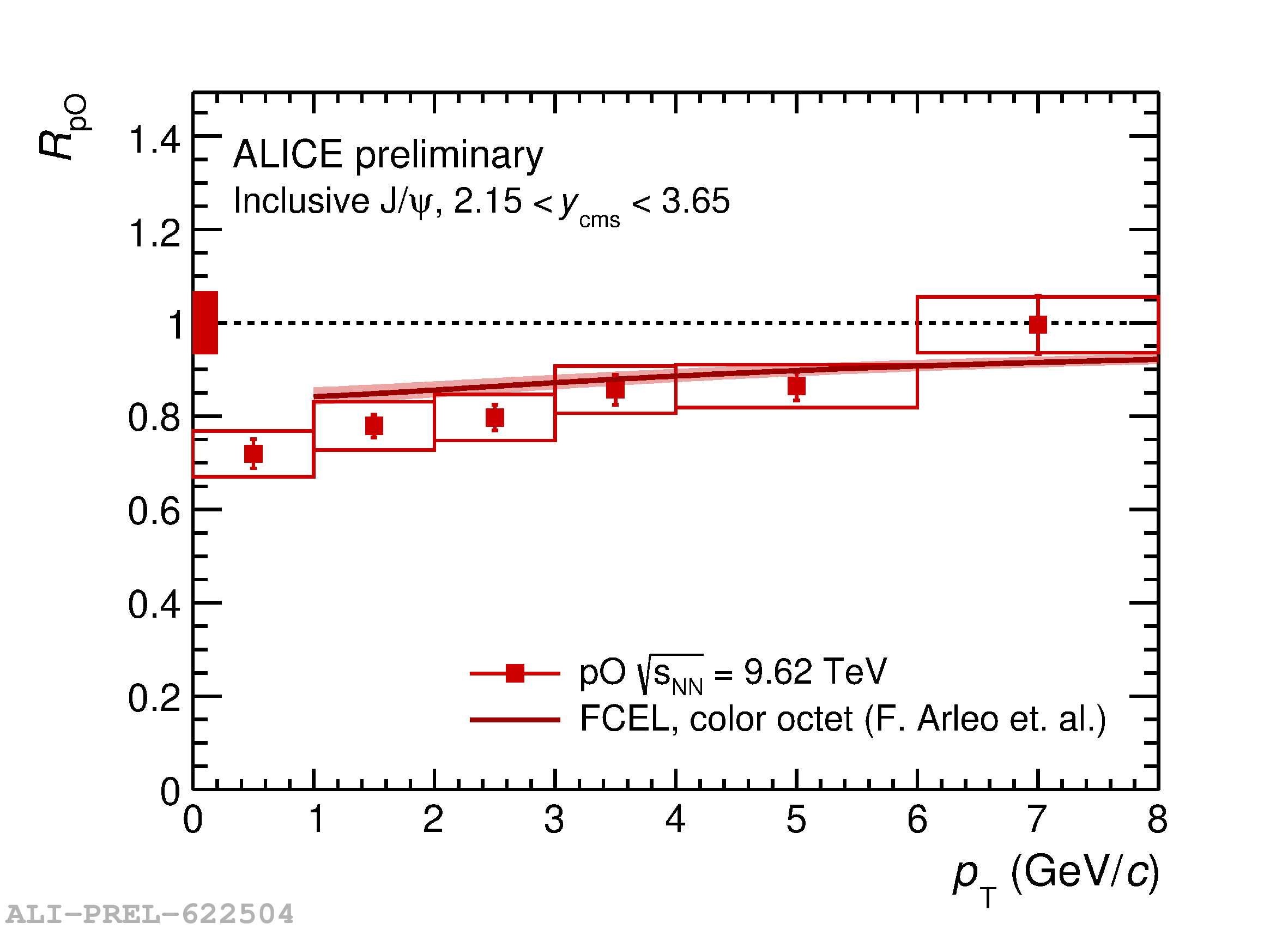}

        \caption{J/$\psi$ nuclear modification factor in pO collisions as a function of $p_{\mathrm{T}}$  at $\sqrt{s_{\mathrm{NN}}} = \text{9.62}$ TeV, compared with FCEL model prediction.}
        \label{fig:RpO_with_model}
    \end{minipage}

\end{figure}

\begin{figure}[H]
\centering
\includegraphics[width=0.8\textwidth]{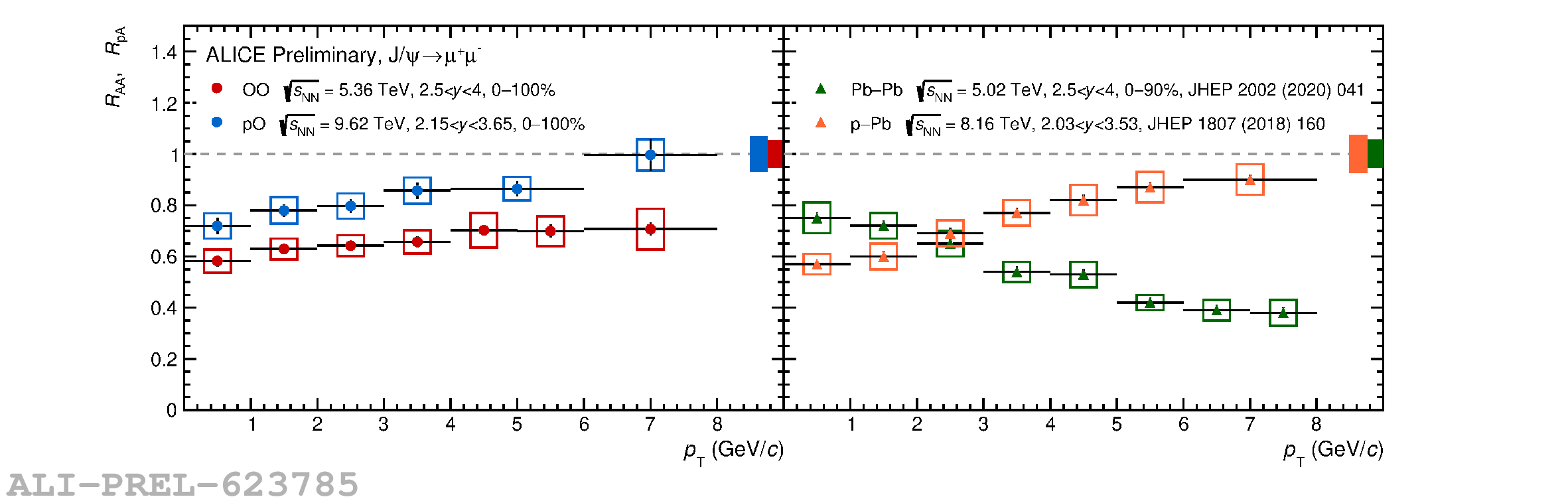}

\caption{J/$\psi$ nuclear modification factor as function of $p_{\mathrm{T}}$ for different collision systems: pO and OO on the left, p--Pb and Pb--Pb on the right.}\label{fig:ROO_RpO_RpPb_RPbPb}

\end{figure}

Figure \ref{fig:ROO_with_model} shows the measured $\mathrm{J}/\mathrm{\psi}$ $R_{\mathrm{OO}}$, which exhibits a suppression of about 40\% at low $p_{\mathrm{T}}$ and only a weak dependence on $p_{\mathrm{T}}$. The result is consistent with transport model prediction \cite{Zhao_2022, Zhou_2020}, which include both cold nuclear matter effects, such as shadowing (EPS09) and the Cronin effect, and hot nuclear matter effects, such as dissociation in the medium and regeneration. The $\mathrm{J}/\mathrm{\psi}$ $R_{\mathrm{pO}}$ is presented in Fig. \ref{fig:RpO_with_model} showing a suppression of approximately 30\% at low $p_{\mathrm{T}}$, followed by a gradual increase towards unity at higher $p_{\mathrm{T}}$. The results are consistent with the FCEL model prediction~\cite{Arleo_2013}. In this model, nuclear effects arise from the constructive interference between the amplitudes for radiation emitted by the incoming and the outgoing partons. The model assumes that the $\mathrm{c}\overline{\mathrm{c}}$ pair is produced in a color octet state and does not include nuclear PDF effects. A complete picture of cold nuclear matter effects is still lacking due to the absence of Op data, which would provide complementary kinematic coverage.
After discussing the OO and pO results separately, it is useful to consider them within a common picture and to compare them with the corresponding measurements in p--Pb and Pb--Pb collisions. Such a comparison allows the observed $p_{\mathrm{T}}$ dependence to be examined across collision systems of different sizes and shapes.
As shown in Fig. \ref{fig:ROO_RpO_RpPb_RPbPb}, the J/$\psi$ $R_{\mathrm{OO}}$ is systematically lower than J/$\psi$ $R_{\mathrm{pO}}$ over the measured $p_{\mathrm{T}}$ range. This behaviour could be expected, since pO collisions are mainly affected by cold nuclear matter effects, whereas in OO collisions additional hot nuclear matter contributions may start to play a role. The comparison between pO and p--Pb 
shows a similar $p_{\mathrm{T}}$ dependence, although the magnitude of the nuclear modification  differs between the two systems. In contrast, the Pb--Pb results exhibit a stronger dependence on $p_{\mathrm{T}}$ than the OO measurements, particularly at low $p_{\mathrm{T}}$, where regeneration is expected to play a more significant role. The difference is not limited to the shape of the $p_{\mathrm{T}}$ dependence. A stronger modification is also observed at high $p_{\mathrm{T}}$, where the $\mathrm{J}/\mathrm{\psi}$  $R_{\mathrm{PbPb}}$ is as low as 0.4, indicating that hot medium effects contribute more in Pb--Pb than in OO collisions. 
The slightly different centrality ranges used for Pb--Pb (0--90\%) and OO (0--100\%) are not expected to significantly affect the comparison, as both samples are dominated by central collisions.

\section{Conclusions}
The first measurements of the $\mathrm{J}/\mathrm{\psi}$ nuclear modification factor in OO and pO collisions by the ALICE collaboration provide a new step towards bridging the results obtained in pp, p–Pb, and Pb–Pb collisions. These measurements provide the first indications of whether hot nuclear matter effects may already play a role in the $\mathrm{J}/\mathrm{\psi}$ production in light-ion collisions, which are systems with different sizes. Although still preliminary, the results
already provide important insights. In particular, the $\mathrm{J}/\mathrm{\psi}$ $R_{\mathrm{pO}}$ can be
described by a model including only cold nuclear matter effects while the $R_{\mathrm{OO}}$ shows a stronger suppression.  This is qualitatively suggesting that hot nuclear matter effects might be at play in OO collisions, which is supported by the good description of the data by models that include them. The comparison between $\mathrm{J}/\mathrm{\psi}$ $R_{\mathrm{OO}}$ and  $\mathrm{J}/\mathrm{\psi}$ $R_{\mathrm{PbPb}}$ further suggests that, although hot medium effects may already be present in OO collisions, the recombination contribution has a smaller magnitude in OO than in Pb--Pb collisions. This may be attributed to the lower abundance of heavy-quark pairs produced in the smaller collision system. In addition, the suppression of $\mathrm{J}/\mathrm{\psi}$ at high $p_{\mathrm{T}}$ is  more pronounced in Pb--Pb than in OO collisions suggesting that the larger system size and higher initial energy density might play a role. The measurements of $R_{\mathrm{pO}}$ and $R_{\mathrm{pPb}}$ complete the picture, showing a similar dependence on the studied kinematic variables, while the overall magnitude differs between the two collision systems. Another interesting future result would be a comparison of $R_{\mathrm{AA}}$ as a function of $N_{\mathrm{part}}$ in OO and Pb--Pb collisions, providing insight into the system-size dependence. Complementary information could be obtained from measurements of the J/$\psi$ $v_{2}$, helping to clarify the role of hot nuclear matter effects in light-ion collisions.

\label{sec1}

\bibliographystyle{elsarticle-num}
\bibliography{sqm2026_template}



\end{document}